# Demand-driven provisioning of Kubernetes-like resources in OSG

*Igor* Sfiligoi[1], *Frank* Würthwein[1], *Jeff* Dost[1], *Brian* Lin[2], and *David* Schultz[2]

[1]University of California San Diego, La Jolla, CA 92093, USA
[2]University of Wisconsin–Madison, Madison, WI 53715, USA

**Abstract.** The OSG-operated Open Science Pool is an HTCondor-based virtual cluster that aggregates resources from compute clusters provided by several organizations. Most of the resources are not owned by OSG, so demand-based dynamic provisioning is important for maximizing usage without incurring excessive waste. OSG has long relied on GlideinWMS for most of its resource provisioning needs but is limited to resources that provide a Grid-compliant Compute Entrypoint. To work around this limitation, the OSG Software Team has developed a glidein container that resource providers could use to directly contribute to the OSPool. The problem of that approach is that it is not demand-driven, relegating it to backfill scenarios only. To address this limitation, a demand-driven direct provisioner of Kubernetes resources has been developed and successfully used on the NRP. The setup still relies on the OSG-maintained backfill container image but automates the provisioning matchmaking and successive requests. That provisioner has also been extended to support Lancium, a green computing cloud provider with a Kubernetes-like proprietary interface. The provisioner logic has been intentionally kept very simple, making this extension a low-cost project. Both NRP and Lancium resources have been provisioned exclusively using this mechanism for many months.

## 1 Introduction

The HTCondor batch workload management system [1,2] has long been used to aggregate resources from many independent resource providers and is the core technology enabling the Open Science Grid (OSG) [3] operated Open Science Pool (OSPool) [4]. HTCondor architecture has very few hard requirements, allowing its services to operate in virtually any environment, e.g., both with and without elevated privileges, and in restricted network environments. Resource provisioning was, however, never a core competency of the HTCondor stack, delegating that aspect to other software providers.

OSPool currently mostly relies on GlideinWMS [5] for its dynamic resource provisioning needs. That said, GlideinWMS specializes in provisioning Grid computing resources, i.e. compute resources managed by independent batch workload management systems behind a Grid-compliant Compute Entrypoint (OSG CE) [6].

In this work we address the provisioning of distributed compute resources that are not managed by a traditional batch system, with a focus on container-based systems. It should be noted that the OSG Software Team has already developed a glidein container image [7] that can be used to contribute resources to the OSPool, which has been successfully used by some resource providers as a backfill solution. Our solution extends that by adding demand-driven provisioning logic on top of that, allowing for OSPool provisioning to work both as backfill and at regular priorities.

## 2 Provisioning Kubernetes-managed resources

Kubernetes [8] is a popular container-based resource management system that is getting significant traction in both on-prem and Cloud environments. In particular, the US National Science Foundation (NSF) has funded several on-prem systems that are at least partially managed by Kubernetes, including the Pacific Research Platform (PRP) [9], Expanse, Voyager and the Prototype National Research Platform (PNRP) [10]. Proper integration of such resources in the OSG ecosystem is thus highly desirable.

An initial Kubernetes provisioner has been developed in PRP [11], but it was mostly focused on supporting a few large communities, e.g., the IceCube and LIGO experiments. In particular, the container image providing the HTCondor worker processes was built starting from a base Operating System (OS) image and had to be customized for each and every target user community.

This work [12] extends that implementation by adding support for the OSG glidein container image. The big advantage of this approach is that the OSG-provided image comes fully pre-configured, avoiding the image customization and maintenance effort needed by the original implementation. The downside of this approach is slightly reduced provisioner flexibility, but we found no showstoppers.

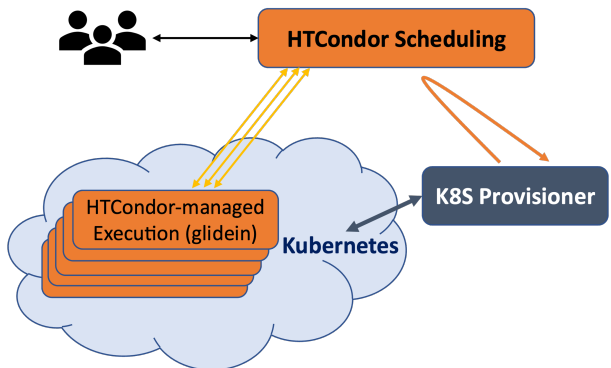

**Fig. 1.** Summary overview of the Kubernetes Provisioner architecture

### 2.1 Provisioning logic

The provisioning logic is based on an asynchronous polling mechanism and is demand-driven. The provisioning process periodically queries both HTCondor and Kubernetes for their relevant state, and if there are HTCondor jobs waiting for resources and no relevant queued Kubernetes requests, additional Kubernetes resources are requested. Once a Kubernetes-managed container starts, it contacts the HTCondor scheduling infrastructure, which in turn sends the user job to be executed, as outlined in Figure 1. It should be noted that conceptually this is very similar to the logic used by GlideinWMS, just optimized for provisioning from a single Kubernetes pool.

The Kubernetes provisioner effectively only manages the up-scaling part of the auto-scaling logic, by queuing more pods as needed. The provisioner never actively removes any pods. All pods are configured with total lifetime and maximum idle time limits, autonomously auto-terminating and thus implicitly managing the down-scaling. This logic avoids race conditions inherent to the asynchronous nature of the logic and was observed to work reasonably well in production environments.

Since HTCondor jobs in the OSPool are virtually never homogeneous, the Kubernetes provisioner groups them by their requirements and requests dedicated Kubernetes resources for each of the groups. This minimizes the waste incurred by the running Kubernetes pods and allows the Kubernetes scheduler to optimally allocate its resources. An overview of the logic is available in Figure 2.

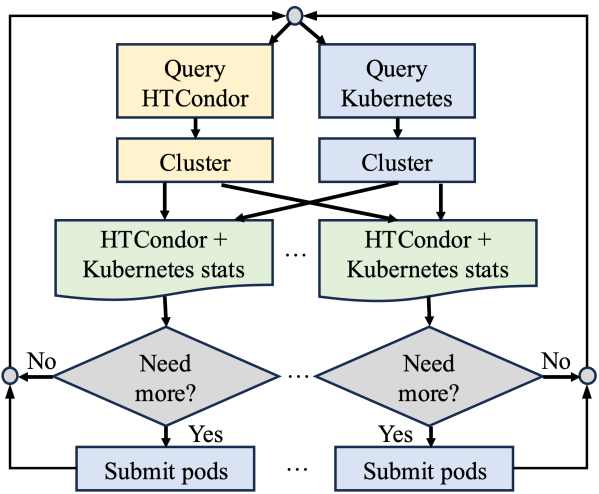

**Fig. 2.** Overview of the clustered provisioning logic

Moreover, not all HTCondor jobs can or want to run on a specific Kubernetes cluster. Neither are all the Kubernetes-managed resources suitable or available to the OSG community. The provisioner thus allows for filtering of HTCondor jobs during its query phase and for adding of additional requirements during Kubernetes pod submission. As with most software, those additional restrictions are controlled though an admin-provided configuration file. A simplified example of such a configuration can be seen in Figure 3.

```
[HTCondor]
additional_requirements= \
   ((DESIRED_Sites is undefined)|| \
     stringListMember("SDSC-PRP",DESIRED_Sites,","))&& \
   ((UNDESIRED_Sites is undefined)|| \
     !stringListMember("SDSC-PRP",UNDESIRED_Sites,","))&& \
   (!isUndefined(ProjectName))&& \
   (!isUndefined(SingularityImage))
[k8s]
node_affinity_dict=^nautilus.io/low-power:true
```

**Fig. 3.** A simplified example configuration file showcasing the provisioning restrictions

### 2.2 Integration with the OSG maintained container image

The container image used by the Kubernetes pods has three main functions:
1. provide the necessary software needed by the user jobs,
2. provide the HTCondor software distribution, and
3. properly configure HTCondor on startup.

By using the OSG-maintained glidein container image, the Kubernetes provisioner software stack does not have to maintain the first two anymore. Most of the configuration is also maintained by the OSG container image but must be dynamically patched at runtime to inject the additional bits and pieces the Kubernetes provisioner relies on. This adds the potential of the two getting out of sync, but so far it has not been a problem yet.

The major requirement of the dynamic patching is the propagation of provisioner-specific attributes used for querying and matchmaking. On top of that, the provisioner also uses a slightly different approach at passing secrets used in authentication, so the HTCondor configuration in the image has to be patched accordingly; while it would have been in principle possible to alter the provisioner secret handling, we decided it was less disruptive to just patch the existing configuration.

### 2.3 Interaction between multiple Kubernetes users

The Kubernetes provisioner is generally oblivious about the activities of other Kubernetes users in the system. The task of sharing the resources between those users is mostly offloaded to the Kubernetes scheduler.

The major knob used by the provisioner is the *priorityClassName* of the submitted Kubernetes pods, which regulates how those pods will be scheduled in relation to other pods in the system. For example, in the National Research Platform (NRP) Nautilus cluster, which contains both the PRP and PNRP nodes, the *opportunistic2* class has the lowest priority number of all the defined classes and allows for preemption, and is thus used for backfill pods. On the other hand, regular-priority pods simply do not explicitly specify any *priorityClassName* at all.

Additionally, the provisioner allows for setting of a quota through the configuration setting *max_submit_pods_per_cluster*. This is especially useful for regular-priority, non-preemptable pods, so a single user does not take over the whole cluster in the absence of cluster-wide quota settings.

## 3 Extending the provisioner to Lancium cloud compute

Lancium is a green computing company, who offered a significant amount of compute resources to OSPool through its cloud computing platform. Unfortunately, at that time its cloud offering used a custom interface, so one of the OSG resource provisioning tools had to be extended to make use of it.

The Lancium cloud interface was container based and provided the usual batch-like actions, e.g., query and submit. After close examination, we determined that the Lancium semantics was rich enough to support our Kubernetes-focused provisioner, even though the syntax was significantly different.

Our Kubernetes-focused provisioner is written in Python language, with Python classes abstracting away the Kubernetes details from the provisioning logic. It was thus relatively easy to implement an alternative class [13] that exposed the same interface but interacted with Lancium instead, due to the fact that we used only the basic Kubernetes capabilities in the original code.

## 4 Summary and conclusions

The Kubernetes-focused provisioner described in this work has allowed the OSG communities, and in particular the OSPool, to successfully and effectively make use of the NSF-funded Kubernetes-managed compute resources, e.g., the NRP. At the time of writing, this provisioner was the only solution available to the OSG Consortium for dynamically provisioning multi-tenant Kubernetes systems.

The work builds on top of the Kubernetes provisioner built as part of the PRP project, but further integrates it with the OSG Software Stack. This both reduces code maintenance and minimizes OSPool job failures due to configuration errors. Nevertheless, other user communities are still supported using the original container image approach.

Additionally, the provisioner has been shown to be easily extendible to support other platforms with a Kubernetes-like interface, due to its minimalistic interface requirements. This was proven by adding support for the Lancium cloud platform, which has delivered a significant amount of resources to the OSPool.

## Acknowledgments

This work was partially funded by the U.S. National Science Foundation (NSF) under grants OAC-2030508, OAC-2112167, OAC-1826967, CNS-1925001, OAC-1841530, CNS-1730158, CNS-2100237 and CNS-2120019.

## References

1. D. Thain, T. Tannenbaum, and M. Livny, Distributed computing in practice: the Condor experience. Concurrency Computat.: Pract. Exper., 17: 323-356. (2005) https://doi.org/10.1002/cpe.938
2. *HTCondor Home Page*. Accessed August 2023. https://htcondor.org/

3. R. Pordes et al. *The open science grid*, J. Phys.: Conf. Ser. 78 012057 (2007)
4. *Open Science Pool*. Accessed August 2023. https://osg-htc.org/services/open_science_pool.html
5. I. Sfiligoi et al., *The Pilot Way to Grid Resources Using glideinWMS*, 2009 WRI World Congress on Computer Science and Information Engineering, pp. 428-432. (2009) https://doi.org/10.1109/CSIE.2009.950
6. B. Bockelman, M. Livny, B. Lin, and F. Prelz, *Principles, technologies, and time: The translational journey of the HTCondor-CE*, J. Comp. Sci. 101213. (2020) https://doi.org/10.1016/j.jocs.2020.101213
7. *Open Science Pool Containers*. Accessed August 2023. https://osg-htc.org/docs/resource-sharing/os-backfill-containers/
8. *Kubernetes*. Accessed August 2023. https://kubernetes.io
9. L. Smarr et al., *The Pacific Research Platform: Making High-Speed Networking a Reality for the Scientist*, PEARC '18: Proceedings of the Practice and Experience on Advanced Research Computing. 29 pp 1–8. (2018) https://doi.org/10.1145/3219104.3219108
10. *NRP*, Accessed August 2023. https://www.sdsc.edu/services/hpc/nrp/index.html
11. I. Sfiligoi, T. DeFanti, and F. Würthwein, *Auto-scaling HTCondor pools using Kubernetes compute resources*, In Practice and Experience in Advanced Research Computing (PEARC '22). Association for Computing Machinery, New York, NY, USA, Article 57, 1–4. (2022) https://doi.org/10.1145/3491418.3535123
12. *GitHub: prp-htcondor-portal*, Accessed August 2023. https://github.com/sfiligoi/prp-htcondor-portal/tree/ospool_pool_base
13. *GitHub: lancium-htcondor-portal*, Accessed August 2023. https://github.com/sfiligoi/lancium-htcondor-portal/tree/main